\documentclass[%
superscriptaddress,
preprint,
 amsmath,amssymb,
prb,
]{revtex4-1}

\usepackage{graphicx}
\usepackage{epstopdf}
\DeclareGraphicsExtensions{.eps,.pdf,.png,.jpg,.gif}
\usepackage{dcolumn}
\usepackage{chemformula}
\usepackage{bm}
\usepackage{hyperref}
\usepackage{color}

\begin{document}


\title{On the Kinetic Energy Density Functional: The Limit of the Density Derivative Order}

\author{Abdulaziz H. Al-Aswad}
\email{abdu@kfupm.edu.sa}
\affiliation{Physis Department, King Fahd University of Petroleum and Minerals, Saudi Arabia.}%

\author{Fahhad H. Alharbi}
\affiliation{Electrical Engineering Department, King Fahd University of Petroleum and Minerals, Saudi Arabia.}
\affiliation{SDAIA-KFUPM Joint Research Center for Artificial Intelligence, Dhahran, 31261, Saudi Arabia}

\date{\today}

\begin{abstract}
Within ``orbital-free'' density functional theory, it is essential to develop general kinetic energy density (KED), denoted as $t(\mathbf{r})$. This is usually done by empirical corrections and enhancements, gradient expansions, machine learning, or axiomatic approaches to find forms that satisfy physical necessities. In all cases, it is crucial to determine the largest spatial density derivative order, $m$ in,  $t(\mathbf{r})$. There have been many efforts to do so, but none have proven general or conclusive and there is no clear guide on how to set $m$. In this work, we found that, by imposing KED finitude, $m=D+1$ for systems of dimension $D$. This is consistent with observations and provides a needed guide for systematically developing more accurate KEDs.
\end{abstract}


\maketitle


\section{\label{Sec1} Introduction}

Currently, density functional theory  (DFT) \cite{H01,J01} in its Kohn-Sham form (KS-DFT) \cite{K01,B01} is the most widely used approach for atomic-scale calculations. The original DFT suggests that the total energy of an electronic system can be calculated via a functional of the electronic density of the system $\rho(\mathbf{r})$. However, finding a general form for the kinetic energy contribution in the form of a density functional (KEDF, $ T[\rho(\mathbf{r})] = \int t(\rho(\mathbf{r})) d\mathbf{r}$ where $t(\rho(\mathbf{r})) \equiv t(\mathbf{r})$ is the kinetic energy density (KED))  is extremely difficult. The origin of this problem preceded  Hohenberg and Kohn’s seminal paper on the theory’s development \cite{H01} and is dated back even to the late 1920s when Fermi \cite{F01} and von Weizs{\"a}cker \cite{V01} derived the KED for two special cases (a uniform electron gas and a singly occupied state, respectively). To overcome this challenge, Kohn and Sham \cite{K01} ``provisionally'' introduced individual electron states (orbitals $\phi_i$) and suggested that the total kinetic energy can be approximated as a sum of the kinetic energies of the non-interacting electrons ($T_s(\phi_i)$) and a ``correction term'' (the exchange-correlation potential $V_{xc}$). Computationally, this converts the problem from a $D$-dimensional problem for a $D$-dimensional system to a highly coupled set of $N\, D$-dimensional equations, where $N$ is the number of electrons in the system.

Although KS-DFT is dominant theory in the field, there have been considerable efforts to restore the original DFT theory and to regain its computational advantages \cite{W01,W02,K02,K03,L01,A01,M01,A03}. 
The goal is to find an accurate and generally applicable KED that does not require the provisional orbitals to be introduced, as described by the commonly used term ``orbital-free'' DFT (OF-DFT). For a comprehensive list of the KEDs that have been proposed so far, we refer the reader to the lists provided by Tran and Wesolowski \cite{W01} and by Liu et al. \cite{L09}. Many approaches were used to develop these KEDs, such as using special solvable models, applying gradient expansion, adopting empirical correction, and many other methods. Machine learning (ML) has also been used to develop KEDs \cite{A01,A03,M02,F02,B02,S01,S02,I01,R02}.

An interesting alternative and ``axiomatic'' approach that can pave the way for a general purpose KED has also been pursued. In this approach, instead of starting from models or adding corrections to existing ones, it identifies all the possible KEDs within a suitable function space that are physically acceptable \cite{G01,A01,M01,L02}. The physical necessities considered are dimensionality, finitude, satisfaction of the virial theorem, and, to a lesser extent, spatial scaling. It is assumed that
\begin{equation}
    \label{KEDFEx}
    t(\mathbf{r}) = \sum_{j} a_j t_j(\mathbf{r}) \,,
\end{equation}
where
\begin{equation}
\label{KEDFf1}
    t_j(\mathbf{r}) \equiv f_j \left( \rho(\mathbf{r}),\nabla \rho(\mathbf{r}), \cdots, \nabla^m \rho(\mathbf{r}) \right) \, .
\end{equation}
Clearly, it is crucial to determine the maximum derivative order $m$. The derivatives are consecutive gradients followed by divergence operators acting on the scalar function of $\rho(\mathbf{r})$.

Determining $m$ has been one of the open questions in the field of DFT, and this is relevant for both the kinetic energy (as in this paper) as well as for the exchange correlation energy \cite{L02,P03,B05}. The commonly employed  solution is to adopt the generalized gradient approximation \cite{P01,P02}, in which functions of the normalized density gradient and Laplacian are used to approximate the desired energy functionals (i.e .,  assuming $m = 2$ for three-dimensional problems). The other extreme is to assume that $m \to \infty$ by using the  Wigner-Kirkwood gradient expansion. This was initially suggested by Kirzhnitz \cite{K04}, who assumed the Thomas-Fermi KED \cite{F03,T01} as the leading term. This was initially applied to slowly varying electron gas. However, it fails for other systems, and the KED diverges for terms that have a derivative greater than or equal to the sixth order \cite{Y01,S03}. As the Wigner-Kirkwood gradient expansion allows only even derivative operations, $m$ could be either 4 or 5 for three-dimensional space. For one-dimensional cases, Baltin \cite{B03,B04} showed that the maximum possible value of $m$ that ensures compatibility with the virial theorem is 1. 

In this work, the upper derivative limit $m$ for a $D$-dimensional system is derived via asymptotic analysis of localized systems. We find that $m = D + 1$, which agrees with the implied values, as discussed in more detail in the analysis and discussion section. Although the analysis focuses on localized systems, the result is also applicable to periodic systems. This should provide a needed guide for systematically developing more accurate KEDs by meta-generalized gradient approximations, physics-guided machine learning implementations, and axiomatically-inspired expansions.

\section{\label{Sec2} Kinetic Energy Density and the Highest Derivative}

The commonly used function space for KEDs is multivariate polynomials either explicitly or implicitly, in which $t_j(\mathbf{r})$ (as in Eq.(\ref{KEDFf1})) is assumed to be a product of powers of electronic density gradients in the form
\begin{equation}
\label{texp1}
t_j(\mathbf{r})\equiv   \rho^{\ell/D} \, \nabla\rho^{n_{1}} \, (\nabla^{2}\rho)^{n_{2}}\; ...\;(\nabla^{m}\rho)^{n_{m}} \, ,
\end{equation}
where $D$ is the number of dimensions and $\ell$ and $n_{m}$ are integers. 
To generally ensure the finitude of the KED, $\nabla^{m}\rho$ must equal 0 at some points to have a finite number of electrons. Thus, $n_m$ must be positive. Eq. \ref{texp1} can be concisely rewritten as
\begin{equation}
\label{texp2}
    t_j(\mathbf{r})\equiv  \rho^{\ell/D} \prod_{k=1}^{m}\;(\nabla^{k}\rho)^{n_{k}} \, .
\end{equation}
This function space encompasses many regularly used KEDs like the local density approximation (LDA), the converging gradient expansion approximation (GEA$n$), the generalized gradient approximation (GGA), and the meta-GGA. GEA$n$ expansion terms take explicitly the same form in Eq. \eqref{texp2}. For LDA, GGA, and meta-GGA, $t(\mathbf{r})$ is assumed to be enhancements to the Thomas-Fermi KED such that $t(\mathbf{r})=t_\text{TF}\,F(\cdot)$ where $t_\text{TF}=c_\text{TF}^{(D)}\rho(\mathbf{r})^{\frac{D+2}{D}}$, $c_\text{TF}^{(D)}$ is the Thomas-Fermi in $D$-dimension, and $F(\cdot)$ is the enhancement function in some variables. In LDA, $F(\rho(\mathbf{r}))$ is a function of the density while for GGA, $F(s)$ is a function of the normalized gradient $s=c_1^{(D)}| \nabla \rho| / \rho^{\frac{D+1}{D}}$ where $c_1^{(D)}$ is a $D$-dependent constant. For meta-GGA, the enhancement function $F(\cdot)$ becomes function of the normalized gradient $s$ and higher normalized derivatives to allow more flexibility. Yet, the most commonly used extensions are based on the normalized Laplacian $q=c_2^{(D)}|\nabla^2 \rho| / \rho^{\frac{D+2}{D}}$. The connection to the multivariate polynomials function space comes from the fact that the enhancement function $F(\cdot)$ is representable by multivariate power series.

Let's define a normalized density $k^\text{th}$-derivative $u_k=|\nabla^k \rho|/ \rho^{\frac{D+k}{D}}$. Then, Eq. \eqref{texp2} can be rewritten as:
\begin{equation}
\label{texp3}
    t_j(\mathbf{r}) =  \rho^{\frac{1}{D} \left( \ell + \sum_{k=1}^{m} n_k(k+D) \right) } \prod_{k=1}^{m}\;u_k^{n_{k}} \, .
\end{equation}
This will be further reduced shortly to form of $=t_\text{TF}\,F(\cdot)$ when the dimensionality is imposed.
Using the dimensionality for Eq. \ref{texp2} where $\rho(\mathbf{r})$ has the dimensions of $L^{-D}$, yields
\begin{equation}
\label{DimT1}
   L^{-D-2} = \, L^{-\ell}\, \prod_{k=1}^{m} (L^{-D-k})^{n_{k}}\, .
\end{equation}
Equating the exponents of both sides of Eq. \ref{DimT1} and solving for $\ell$ gives
\begin{equation}
\label{lExc}
 \ell = \,D + 2-\sum_{k=1}^{m} (D+k)n_{k}  \,.
\end{equation}
So, Eq. \eqref{texp3} becomes
\begin{equation}
\label{texp4}
    t_j(\mathbf{r}) =  \rho^\frac{D+2}{D} \prod_{k=1}^{m}\;u_k^{n_{k}} \, .
\end{equation}
By applying the above form in Eq. \eqref{KEDFEx}, it becomes
\begin{equation}
    \label{KEDFEx2}
    t(\mathbf{r}) = \sum_{j} \left[ a_j \rho^\frac{D+2}{D} \prod_{k=1}^{m}\;u_k^{n_{k}^{(j)}} \right] = t_\text{TF} \sum_{j} \left[ \dfrac{a_j}{c_\text{TF}^{(D)}} \prod_{k=1}^{m}\;u_k^{n_{k}^{(j)}} \right]  \,,
\end{equation}
where $n_{k}^{(j)}$ is the power of the $k^\text{th}$ density derivative for the $j^{th}$ KED term. Clearly, this encompasses the regular enhancement functions in GGA and meta-GGA.

For localized systems, the density $\rho(\mathbf{r})$ and its derivatives $\nabla^{k}\rho(\mathbf{r})$ can be generally represented by
 \begin{equation}
 \label{Eq8}
  \begin{split}
\rho(\mathbf{r})=f_0(\mathbf{r}) \,e^{-b r} \, ,
\\
\nabla^{k}\rho(\mathbf{r})= f_k(\mathbf{r}) \,e^{-b r} \, ,
 \end{split}
\end{equation}
where $b\in\mathbb{R}_{>0}$ and $f_0(\mathbf{r})$ is a function of \textbf{r} that is exponentially bounded and that grows slower than $e^{b r}$ (i.e. $f_0(\mathbf{r})<M e^{b r}$, where $M$ is a finite positive number).

This condition is a physical and mathematical necessity, as $\rho(\mathbf{r})$ must be real, non-negative, and Riemann integrable such that $\int \rho(\mathbf{r}) \, d\mathbf{r}=N$ where $N$ is a finite number of electrons. Such a condition is also satisfied by any density that approaches zero faster than any exponential such that $\rho(\mathbf{r}) \propto e^{-\alpha r^\zeta}$ where $\alpha\in\mathbb{R}_{>0}$ and $\zeta>1$. For example, $\zeta=2$ for harmonic oscillators. For these densities, $b$ can be any real positive number. In some cases, $f_0(\mathbf{r})$ contains powers of $r$, which generally grow slower than the decaying exponential (i.e. $e^{-br}$).

For molecules \cite{G03,S06,H04,L08,H03}, the electronic density is approximated asymptotically as $r \to \infty$ by 
\begin{equation}
    \label{AsymN}
    \rho(\mathbf{r}) \to r^\beta e^{-2 \sqrt{2 \left(I+v_\text{eff}(\infty) \right)} r}
\end{equation}
where $I$ is the ionization energy, $\beta=1/\sqrt{2\left(I+v_\text{eff}(\infty) \right)}-1$, and $v_\text{eff}(\infty)$ is the effective potential when $r \to \infty$. Similar exponential forms are expected for any Coulombic system.

Using the forms of $\rho(\mathbf{r})$ and its derivatives $\nabla^{k}\rho(\mathbf{r})$, as expressed in Eq. \ref{Eq8}, produces the general KED form (Eq. \ref{texp2}) of
\begin{equation}
 \lim_{r \to \infty}t_j(\mathbf{r}) \cong \left[   f_0(\mathbf{r})^{\ell/D} \prod_{k=1}^{m}\;f_k(\mathbf{r})^{n_{k}} \right] e^{-q\,b\,r} \, ,
\end{equation}
where $q=\ell/D+\sum_{k=1}^{m} n_k$. In order for $t_j(\mathbf{r})$ to be finite, $q$ has to be greater than zero (as $b > 0$), and hence
 \begin{equation}
 \label{Eq11}
 - \ell < \sum_{k=1}^{m} n_k\, D \, .
\end{equation}
Substituting the value of $\ell$ from Eq. \ref{lExc} into Eq. \ref{Eq11} and solving for $D$ yields
\begin{equation}
 \sum_{k=1}^{m} k\,n_k \, < D + 2 \,.
\end{equation}
This equation relates the dimension $D$ of the problem and the highest spatial derivative order $m$. Therefore, we can use it to determine the maximum possible $m$ for a given dimension $D$. To maximize $m$, we should set all $n_k$ for $k<m$ to be zero. This results in

\begin{equation}
\label{Eq13}
 m\,n_m < D+2 \,.
\end{equation}
The maximim $m$ is obtained by having the lowest finite $n_m$ (i.e., 1). Therefore,
\begin{equation}
\label{mMax}
  m = D+1 \, .
\end{equation}
This is an important result because it gives the maximum possible spatial derivative of the density for asymptotically finite KEDs in $D$-dimension electronic systems. 

Although our analysis focused on localized systems, it is also applicable to periodic systems. In this case, $b \to 0$ and $\rho(\mathbf{r})$ and its derivatives  $\nabla^{k}\rho(\mathbf{r})$ (Eq. \ref{Eq8}) become periodic. This effectively amounts to changing all the ``less than'' signs in Eqs. \ref{Eq11}-\ref{Eq13} to ``less than or equal to'' signs. This results in $m=D+2$, which causes the KED to diverge for localized systems. If KED universality is to be prioritized \cite{V02,P03,G02,P04}, this is not acceptable, and hence $m=D+1$ is the limit for both localized and periodic systems.

\section{\label{Sec3} Analysis and Discussion}

Table \ref{tab:table1} lists the highest spatial derivative order $m$ as obtained in this work and the implied values found in the literature for low-dimensional spaces. For one-dimensional cases, asymptotic analysis reveals that $m = 2$. Nevertheless, Baltin \cite{B03,B04} suggested that $m = 1$ should be used to ensure compatibility with the virial theorem. However, Baltin’s approach was challenged recently by Luo and Trickey \cite{L02}. By considering $m = 2$, only one additional term is added to the list proposed in earlier work \cite{A01}: $d^2 \rho /dx^2$. However, for a well-behaving density, $\int_{-\infty}^\infty d^2 \rho /dx^2 dx \, \equiv 0$. Thus, the additional term for one-dimensional cases does not contribute globally. This fact has deeper implications \cite{A02,E01}.

\begin{table} [ht]
\centering
\caption{\label{tab:table1} Highest spatial derivative order m as obtained in this work and the implied values found in the literature for low-dimensional spaces.}
\begin{tabular}{ccc}
\\
\hline
\hline
$D$ & $m$ & $m$ in literature  \\
\hline
1 & 2 & 1 \cite{B03,B04} \\
2 & 3 &  \\
3 & 4 & $<6$ \cite{Y01} \\
\hline
\hline
\end{tabular}
\end{table}

For three-dimensional problems, the analysis suggests that $m = 4$. This agrees with the known fact that full density gradient expansion causes KEDs to diverge for terms that have a derivative greater than or equal to the sixth order \cite{Y01,S03}. 
This implies that $m$ should be 4, as the Wigner-Kirkwood gradient expansion allows only even derivative operations.

Recently, there have been a lot of efforts to develop accurate KEDs using the enhancement function form $t(\mathbf{r})=t_\text{TF}\,F(\cdot)$. $F(\cdot)$ is developed either theoretically for specific potentials or empirically using statistical learning means. In both cases, most forms assume that the highest derivative order is the Laplacian. However, this works suggested that we may need to extend the allowed derivative orders to $D+1$ ($m=4$ for 3-dimensional problems).

\section{\label{Sec4} Conclusion}

In this study, the highest spatial derivative order $m$ for the KED ($t(\mathbf{r})$) for $D$-dimensional electronic systems was determined mathematically by considering only dimensionality and finitude. We found that $m=D+1$, which agrees with the results from the literature for systems of one and three dimensions. This was previously unknown for $D$-dimensional electronic systems with $D > 1$ and has been one of the open questions in the field of DFT for both the kinetic energy and the exchange-correlation energy. Thus, this paper provides an important finding, which may lead to more precise KEDs that consider proper physical insights, without which the contributions of spatial derivatives of the density $\rho(\mathbf{r})$ are limited and highly empirical. Although our analysis focused on localized systems, it is also applicable to periodic systems. Considering KED universality, $m=D+1$ is the limit for both localized and periodic systems. As aforementioned, this should provide a needed guide for systematically developing more accurate KEDs by meta-GGA, physics-guided machine learning implementations, and axiomatically-inspired expansions.


 
\bibliographystyle{apsrev4-1}
\bibliography{OFDFT}

\end{document}